\documentclass[sigconf]{acmart}
\usepackage{amsmath} 
\usepackage{todonotes}
\usepackage{multirow}
\usepackage{hhline}

\usepackage{makecell}

\AtBeginDocument{%
  \providecommand\BibTeX{{%
    \normalfont B\kern-0.5em{\scshape i\kern-0.25em b}\kern-0.8em\TeX}}}

\copyrightyear{2023}
\acmYear{2023}
\setcopyright{acmlicensed}
\acmConference[WWW '23 Companion]{Companion Proceedings of the ACM Web Conference 2023}{April 30-May 4, 2023}{Austin, TX, USA}
\acmBooktitle{Companion Proceedings of the ACM Web Conference 2023 (WWW '23 Companion), April 30-May 4, 2023, Austin, TX, USA}
\acmPrice{15.00}
\acmDOI{10.1145/3543873.3587657}
\acmISBN{978-1-4503-9419-2/23/04}
\begin{document}

\title{Contextual Response Interpretation for Automated Structured Interviews: A Case Study in Market Research}

\author{Harshita Sahijwani}
\affiliation{%
  \institution{Emory University}
  % \streetaddress{400 Dowman Drive}
  % \city{Atlanta}
  % \state{Georgia}
  \country{USA}
  % \postcode{30307}
  }
\email{hsahijw@emory.edu}

\author{Kaustubh Dhole}
\affiliation{%
  \institution{Emory University}
  \country{USA}
  }
\email{kdhole@emory.edu}

\author{Ankur Purwar}
\affiliation{%
  \institution{Procter \& Gamble}
  \country{Singapore}
  }
\email{purwar.a@pg.com}

\author{Venugopal Vasudevan}
\affiliation{%
  \institution{Procter \& Gamble}
  \country{USA}
  }
\email{vasudevan.v@pg.com}

\author{Eugene Agichtein}
\affiliation{%
  \institution{Emory University}
  \country{USA}
  }
\email{yagicht@emory.edu}

\renewcommand{\shortauthors}{Sahijwani, et al.}
\newcommand{\FORM}{{\it Web-based Questionnaire~}}
\newcommand{\CHAT}{{\it Conversational Interface~}}
\newcommand{\TASK}{response interpretation~}

\begin{abstract}
Structured interviews are used in many settings, importantly in market research on topics such as brand perception, customer habits, or preferences, which are critical to product development, marketing, and e-commerce at large. Such interviews generally consist of a series of questions that are asked to a participant.
These interviews are typically conducted by skilled interviewers, who interpret the responses from the participants and can adapt the interview accordingly. 
Using automated conversational agents to conduct such interviews would enable reaching a much larger and potentially more diverse group of participants than currently possible. 
However, the technical challenges involved in building such a conversational system are relatively unexplored. 
To learn more about these challenges, we convert a market research multiple-choice questionnaire to a conversational format and conduct a user study.
We address the key task of conducting structured interviews, namely interpreting the participant's response, for example, by matching it to one or more predefined options.
Our findings can be applied to improve response interpretation for the information elicitation phase of conversational recommender systems.

\end{abstract}



\maketitle
\vspace{-0.2cm}
\section{Introduction}
\label{sec: introduction}
Information elicitation conversations, such as when a sales agent tries to understand their customer's preferences or a medical professional asks about a patient's history, often begin with a routine set of questions. In e-commerce, market research professionals and companies conduct many such surveys each year, often multiple times, before developing, updating, or launching new products - to collect critical data on customer preferences, interests, and awareness, among other topics.

In structured interviews, an interviewer asks a predetermined set of questions conversationally, adapting them to the user's responses and behavior.  While extremely informative and a de-facto standard in market research (e.g., via focus groups), these studies are limited in scale to a small number of participants and are time-consuming and expensive to conduct. 

To expand the reach of such studies,  online static multiple-choice questionnaires or surveys are used.
However, such online questionnaires have some disadvantages. They need to be shorter than interviews to avoid "respondent fatigue" \citep{bryman2016social}.
There is also a greater risk of missing data because of a lack of probing or supervision. Also, it is difficult to ask open-ended questions \citep{bryman2016social}.
Conversational systems that can conduct structured interviews can thus potentially be more effective tools for preference elicitation.
\begin{figure}
    \centering
    \includegraphics[width=\columnwidth]{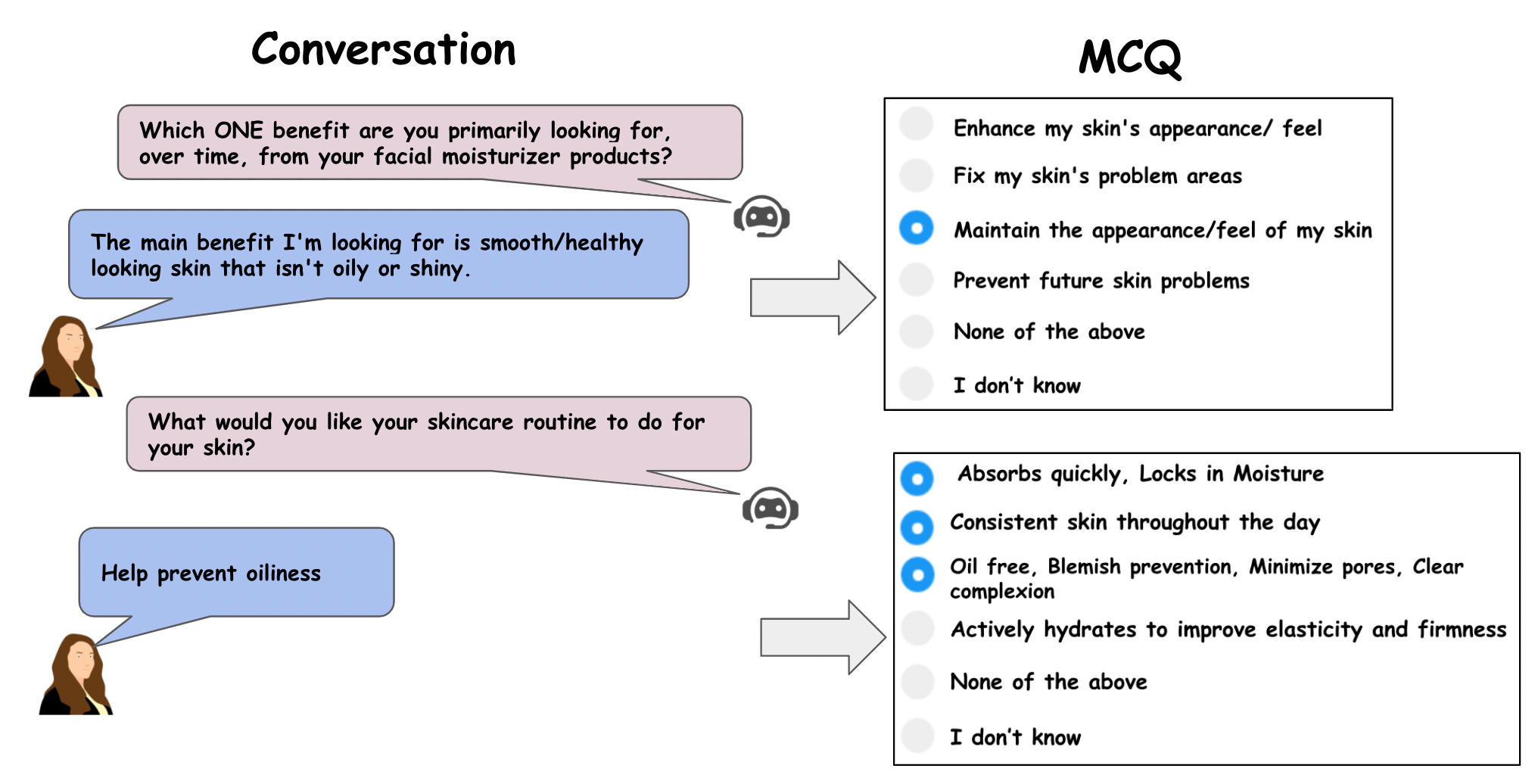}
    \caption{The user's conversational responses should be mapped to the correct answer option(s).}
    \label{fig:mainfig}
    \vspace{-0.5cm}
\end{figure}
Such a system would, given a structured interview provided by a domain expert, converse with the participant to elicit responses to a series of questions.
Ideally, it should also be able to ask clarification questions, prime the user with possible answers, and reorder and skip questions based on the user's responses.
An essential requirement for such an agent to be effective is the ability to interpret the responses, often by matching them to a previously defined set of options.

As a first step towards building a conversational system for conducting structured interviews, we investigate the trade-offs of conducting a structured interview via an automated conversational agent vs. the traditional, static, multiple-choice web-based questionnaire.
To this end, we conduct a large online user study where a questionnaire with choices for each question is presented in both a conversational interface and as a static multiple-choice questionnaire. 
The questionnaire was provided by a reputed Personal Care products company's marketing team. The company has a wide range of products for skin care, which target specific skin conditions.
Market research and brand awareness are critical for ensuring that their products meet their consumers' needs and that they can find the right product.

We then address the \TASK problem for this setting, i.e., given a structured interview in the form of a list of questions and the set of possible answers (options) for each question, the model needs to infer the options with which the user's response matches. 
For the related problem of intent classification for goal-oriented and open-domain conversational agents, prior work achieves good results by jointly training large language models on intent classification and slot-filling tasks. 
However, in a system-initiative conversation where the user is asked open-ended questions about their preferences, intent classification is challenging because 1) interview questions often elicit descriptive answers as opposed to names of entities of an expected type, and 2) it is expensive to collect conversational data for supervised learning.
We investigate three approaches for using contextual information for response interpretation: 1) using historical probability distribution over the answer options, 2) using previous conversation context, and 3) using external knowledge.

Our research questions are RQ1) Does the change in interface, and the absence of options lead to more informative responses? RQ2) What types of questions would benefit from an open-ended conversational interface? And RQ3) How can we address the response interpretation problem (defined below) for this setting?

% \todo[inline]{Add pararagraph here overviewing the rest of the paper }

\begin{figure}[h!]
\centering
\noindent\begin{tabular}{l l}
\hline
 \textbf{Setting:} & Structured interview conducted by a conversational  \\   & agent with a user \\
 \textbf{Given:} & A conversation consisting of system utterances\\
                & (in the form of questions)\\
                & $s_1$... $s_{n-2}$, $s_{n-1}$, $s_{n}$,\\ 
                & and user responses \\  & $u_1$... $u_{n-2}$, $u_{n-1}$, $u_{n}$,\\
                & and a set of possible answers to $s_{i}$ given by \\ & $A(q=s_i) = a_{i,1},...,a_{i,m}$  \\
 \textbf{Problem:} & At conversation turn $i$, match $u_{i}$ to a subset $M_i$ \\ & of possible answer options ${A(q=s_i)}$ that represents \\
 & user intent\\
 \hline
 \\

\end{tabular}
\label{fig:im-problem}
\vspace{-0.5cm}
\caption*{Response Interpretation Problem Definition}
\end{figure}

\section{Related Work}
\label{sec: related work}
There has been extensive prior work on closely related problems like intent prediction and slot-filling for conversational systems \citep{qin2021co, wen2017jointly, weld2021survey}, dialog representation \citep{mehri2019pretraining, ortega2017neural}, knowledge grounded language models \citep{zhang2021knowledge}, and domain-specific language models \citep{beltagy2019scibert}. 

Open-domain and domain-specific conversational agents usually have a predefined set of intents and slot values that they can identify and process. Existing intent classifiers apply a variety of approaches like transformer-based models \citep{qin2021co}, hierarchical text classification \citep{wen2017jointly}, and knowledge-guided pattern matching \citep{weld2021survey} to map user utterance to the relevant intent. 
% Both open-domain and domain-specific conversational agents usually have a predefined set of intents and slot values that they are capable of identifying and parsing. Existing intent classifiers apply a variety of approaches from transformer-based models \citep{qin2021co}, hierarchical text classification \citep{wen2017jointly} to knowledge-guided pattern matching \citep{weld2021survey} to map user utterance to the relevant intent. 
However, these methods rely on the availability of extensive training data and the intents and slots being limited in number. In the structured interview setting, users often give long descriptive answers to open-ended questions, which makes it hard to apply these intent classification models.

Reading comprehension tasks that require answering multiple-choice questions based on some given context are also closely related to our task.
\citet{luo2020spoken} propose a BERT-based framework for handling multiple-choice questionnaires focused on reference passages.
\citep{ohsugi2019simple, huang2018flowqa} address the problems of history selection and dialog representation for conversational reading comprehension.
However, answers in reading comprehension tasks are generally factual and precise as opposed to ones in structured interviews. The challenges involved in training models for this task are different.

Language Models pre-trained on dialog\citep{wu-etal-2020-tod, zhang-etal-2020-dialogpt} are also relevant to our work.
TOD-BERT~\cite{wu-etal-2020-tod}, after being pre-trained on nine human-human and multi-turn task-oriented dialogue datasets, outperformed strong baselines like BERT on four downstream task-oriented dialogue applications.
We use TOD-BERT in our experiments to study the advantages of dialog pre-training for our task.

External knowledge bases and knowledge graphs have been incorporated in many approaches for NLP and IR tasks to yield promising results ~\cite{yasunaga-etal-2021-qa, annervaz2018learning, lin2019kagnet, kim2021model, liu2021heterogeneous}.
Most of these approaches rely on the existence of a knowledge graph with relevant information. 
Domain-specific models like SciBERT\citep{beltagy2019scibert} and BioBERT \citep{lee2020biobert} have shown that downstream tasks can greatly benefit from models pre-trained on in-domain data.
Although our data is domain-specific, there isn't a pre-trained model or knowledge graph tailored for our setting. Therefore, we use ConceptNet neighbors of terms in conversations to experiment with the effects of incorporating external knowledge.

\section{Data Collection}
\subsection{User Study}
\label{sec: conv data}
We conducted a user study with 139 participants to compare the informativeness and other characteristics of \CHAT responses with \FORM responses.  
We used a questionnaire provided by domain experts from a reputed company, as described in \S \ref{sec: introduction}. 
% Table \ref{tab:user study data} shows other details about the study.
It contains 25 multiple-choice questions about the client's lifestyle, skin and hair care routines, and preferences. 
The questionnaire contains 12 single-option questions (the user can select exactly one option) and 13 multi-option questions (the user can select multiple options). 
The user study consists of 2 phases. In the first phase, the participants interact with a text-based conversational agent that asks a question from the questionnaire, responds to the user's free-form answer with an acknowledgment (``Ok'', ``Alright'' or ``I see''), and then proceeds to ask the next question. The participants are then asked to fill out an online web-based survey with the same questions, but this time with options to choose from. They were shown their conversational response to the question and asked to pick the options that matched it.
In addition to the responses from the questionnaire, the participants could also choose from two additional options, ``None of the above'' and ``I don't know''.

For our experiments, we only use single-option questions.

\subsection{Response Interpretation Data}
\label{sec: classification data}
We model the response interpretation task as a binary classification problem. That is, given a <conversational response, answer option> pair, the model predicts the probability that they are semantically equivalent.
We use the data from the user study in \S \ref{sec: conv data} as a source of ground truth for <conversational response, answer option> pairs.
We split conversations among the train, validation and test sets in a 60:20:20 ratio.
We construct a labeled dataset of <conversational response, answer option> pairs from conversations in the train set to train our binary classification models. The <conversational response, answer option> pairs from \S \ref{sec: conv data} are used as positive examples.
We add an equal number of randomly selected negative examples. 
The model is trained on 22865 samples and validated on 7724 samples. 
It is then evaluated on the holdout set of 20\% of the conversations.

\section{Methods}
This section describes the different methods we use for response prediction.
\subsection{Using Probabilistic Models Learned from Historical Data}
We use purely probabilistic models, which do not consider response text, as baselines.
\subsubsection{Context-Less: Using Prior Probability Distributions}
In this method, we infer the prior probability distribution over the options for each question using the training data. 
We infer the probability of an answer option $a_{j,k} \in A(s_j)$ being the match for question $s_j$ as follows:
\begin{equation}
    P(M_j = \{a_{j,k}\}) = \frac{\mathcal{N}(a_{j,k})}{\sum_{i=1}^{m} \mathcal{N}(a_{j,i})}
\end{equation}
where $\mathcal{N}(a_{j,i})$ represents the number of times $a_{j,i}$ is observed as the matching choice $M_j$ for $s_j$ in the training data.
The model prediction is therefore $a_{j,k}$, where $k = \operatorname*{argmax}_{x} P(M_j = \{a_{j,x}\})$.

\subsubsection{Contextual: Probability Distribution Conditioned on One Previous Response}
\label{sec: cond prob}
In this method, we use a conditional probability distribution.
Given that $a_{i} \in A(s_i)$ was the selected option for $s_i$, the probability that $a_{j,k} \in A(s_j)$ will be selected for $s_j$, where $i<j$ is given by
\begin{equation}
P(M_j = \{a_{j,k}\} | M_i = \{a_{i}\}) = \frac{P(M_j = \{a_{j,k}\} \text{ and } M_i = \{a_{i}\})}{P (M_i = \{a_{i}\})}
\end{equation}
Intuitively, if the answer to $s_i$ provides some information about the answer to $s_j$, then $H(M_j) > H(M_j|M_i)$,
% \begin{equation}
%     H(M_i) > H(M_i|M_j)
% \end{equation}
where $H(x)$ is the entropy of the probability distribution over the values of random variable x.
\begin{equation}
   H(x) = -\sum_{i=1}^n p(x_i) log_2 p(x_i)
\end{equation}
% Our hypothesis is that the answer options matching the user's conversational responses for different questions in the interview are not independent of each other. Each answer contains information that can be used to infer subsequent answers more accurately.
For example, we observe in our dataset that if the user's response for the question ``After applying a facial moisturizer, how do you want your skin to feel?'' is known, the entropy of probability distribution over the options for ``What type of weather do you usually live in?'' is much lower than the prior.
We find the conditional probability distribution with the lowest entropy as follows:
    \begin{equation}
    \operatorname*{argmin}_{i} H(M_j|M_i)
\end{equation}
The model prediction is therefore $a_{j,k}$ where $k=\operatorname*{argmax}_{x} P(M_j = \{a_{j,x}\} | M_i = \{a_{i}\})$.

\subsection{Fine-tuning Pre-Trained Language Models}
\label{sec: finetuning}
In this approach, we treat response matching as a binary classification task. Given a <conversational response, answer option> pair, we train the model to output a score that indicates their semantic similarity.
The final prediction is the option with the highest score.

\subsubsection{Fine-Tuned BERT Classifier}
\label{sec: bert}

In this method, we fine-tune BERT~\cite{devlin-etal-2019-bert} to output a score of either 1 (when conversational response and answer option match) or 0 (when conversational response and answer option don't match) when given the conversational response and answer option as input.
We employ a linear layer on top of the [CLS] token for classification. 

We predict the semantic similarity score of a user response $u_j$ with all the possible answer options for the question $s_j$ as follows:
% Let $q$ be a question to be posed to the user with option choices given by the function $A(q) = {a_{1} ,..., a_{m}}$ and let $u_n$ be the user response for the same. Then, we compute all options $Y$ greater than a threshold $t \in [0,1]$ in the following fashion:
\begin{equation}
S_{j,k} = BERT([CLS] \| u_j \| [SEP] \| a_{j,k}) \; \; \forall a_{j,k} \in A(q=s_j)
\end{equation}
The model prediction is $a_{j,k}$, where $k = \operatorname*{argmax}_{x} S_{j,x}$.
% \begin{equation}
%   j = \operatorname*{argmax}_{x} s_{j}
% \end{equation}
% \begin{equation*}
%     Y = \{j \; | \; s_{j} > t \}
% \end{equation*}
%Write about what loss is backpropagated
\subsubsection{Incorporating Conversation Context}
\label{sec: bert with context}
We include conversation context in the model input in addition to the conversational response.
% To measure if previous conversational history is useful to better identify the correct choice, we include previous conversation turns in the model input. 
% To keep the conversational nature of the text intact, 
We append each conversational utterance with either a ``[SYS]'' or a ``[USR]'' token depending on whether it is a system or a user utterance. 
Let $t_j$ represent the concatenation of the $j^{th}$ system and user utterances.
\begin{equation*}
t_j = [SYS] \| s_j \| [USR] \| u_j
\end{equation*}
We experiment with three settings:
\begin{itemize}
  \item Context of the current turn $j$:
  \begin{equation*}
S_{j,k} = BERT([CLS] \| t_j \| [SEP] \| a_{j,k}) \; \; \forall a_{j,k} \in A(q=s_j)
\end{equation*}
% \begin{equation*}
%     Y = \{j \; | \; s_{j} > t \}
% \end{equation*}
  \item Context of 1-previous turn:
  \begin{equation*}
      S_{j,k} = BERT([CLS] \| t_{j-1} \| t_j \| [SEP] \| a_{j,k}) \; \; \forall a_{j,k} \in A(q=s_j)
  \end{equation*}
%   \begin{equation*}
%     Y = \{j \; | \; s_{j} > t \}
% \end{equation*}
  \item Context of 2-previous turns:
  \begin{equation*}
      S_{j,k} = BERT([CLS] \| t_{j-2} \| t_{j-1} \| t_j \| [SEP] \| a_{j,k}) \; \; \forall a_{j,k} \in A(q=s_j)
  \end{equation*}
%   \begin{equation*}
%     Y = \{j \; | \; s_{j} > t \}
% \end{equation*}
\end{itemize}
The model prediction is $a_{j,k}$, where $k = \operatorname*{argmax}_{x} S_{j,x}$.

\subsubsection{Incorporating Dialog Pre-training}
\label{sec: todbert}
% The BERT model~\cite{devlin-etal-2019-bert} in our previous experiment has been pretrained over free-flowing text and fine-tuned over dialog data. 
We hypothesize that a model pre-trained on dialog tasks would perform better than a generic pre-trained language model in our conversational setting.
In this approach, fine-tune TOD-BERT instead of BERT. TOD-BERT has the same architecture as BERT but has been pre-trained on various dialog tasks.

\subsubsection{Incorporating External Knowledge}
\label{sec: cnnet}
BERT often does not capture the semantic relatedness of domain-specific terms.
To bridge the vocabulary gap between the user responses and questionnaire answer options, we concatenate one-hop neighbors from ConceptNet \footnote{\em \url{https://conceptnet.io/}} of all the terms in the user input to the user input. We exclude infrequent neighbors to avoid adding noise to our input text.

\section{Experimental Setting}
We use 5-fold cross-validation for our experiments. We treat each fold as the test set one by one and use the other folds as train and validation. We report the average of results from all test folds.

\subsection{Models Compared}
% \todo[inline]{Need intro: explain Why were these models chosen? Do they complement each other/represent different approaches??}
% We use a naive probabilistic model described in \S \ref{sec: cond prob} as the baseline.  
% \textbf{Some linking text here}
\begin{itemize}
    \item Probabilistic Baseline: We use the conditional probability-based model described in \S \ref{sec: cond prob} as the baseline.
    \item BERT: We fine-tuned bert-base-uncased\footnote{\em \url{https://github.com/google-research/bert/blob/master/README.md}} on our dataset of <conversational response, answer option> pairs (\S \ref{sec: bert}). We experiment with different lengths of conversation context. Results are reported for the best version, which only considers the current conversation turn.
    \item TOD-BERT: We also tried a BERT model pre-trained on conversational data. Results are reported for TOD-BERT (described in \S \ref{sec: todbert}) fine-tuned on our task with 2 previous turns of context.
    \item BERT-CNNet: Since our dataset is domain-specific and has a different vocabulary than BERT's pre-training data, we also experiment with augmenting input to BERT with domain-specific keywords. Again, results are reported for the best version that only considers the current conversation turn. (\S \ref{sec: cnnet})
\end{itemize}

\subsection{Evaluation Metric}
For this paper, we train and evaluate our models on single-option questions. 
Therefore, we use accuracy as the evaluation metric, which we define as the fraction of test questions where the model assigns the highest score to the true answer option based on the ground truth data  described in \S \ref{sec: classification data}.

\subsection{Human Annotation}
\label{sec: agreement}
We observed that in the user study, in the \FORM, the participants often selected options that they hadn't implied in their \CHAT responses.
% While this means that the dataset is noisy for response matching, it provides insight into the difference in \FORM and \CHAT responses.
% To verify that the MCQs responses from the user study matched the corresponding conversation responses, we conducted extensive human annotation. 
 To measure how difficult response interpretation is for humans, we recruited annotators from MTurk who were familiar with and interested in the domain. We asked them to choose the most appropriate option for each question, given the chat responses from the original user study participant. Four different workers annotated each question for a sample of 27 conversations. 
% Table \ref{tab:iaa} shows the inter-annotator agreement among the MTurk workers, and between the Mturk workers' majority vote and the original user's choices.
We use Fleiss Kappa \citep{fleiss1971measuring} to measure inter-annotator agreement. The average agreement is 0.46, which indicates moderate agreement. However, it varied significantly across different questions, as Table \ref{tab: questionwise_results} shows.
The average agreement between the MTurkers and original respondents is 0.44, which is also moderate.

% We find agreement is highest for the questions which are general in nature ( not domain specific), have a small number of options that are also short and non-ambiguous (e.g. store names v/s descriptions of skin texture). 

\begin{table}[h]
    \centering
    \small
    \caption{Main Results: Accuracy on Single-Option Questions}
    \begin{tabular}{llrlr}
    Model & \multicolumn{2}{r}{Overall}   & \multicolumn{2}{r}{On High-$\kappa$ Questions} \\
    & Accuracy & Std                  & Accuracy & Std                \\
    \hline
    Prob. Baseline &           0.51  & 0.02 & 0.53  & 0.02 \\
    BERT &          \textbf{0.64 (+24.0 \%)} & 0.04 & \textbf{0.71 (+34 \%)} & 0.04 \\
    TOD-BERT &           0.55 (+7.6 \%) & 0.04 & 0.63 (+18.8 \%) & 0.03 \\
    BERT-CNNET &          \textbf{0.62 (+20.9 \%)} & 0.02 & \textbf{0.68 (+28.3 \%)} & 0.05 \\
    \end{tabular}
    \label{tab:main_results_accuracy}
\end{table}

\section{Results and Discussion}
We first report the main results of different methods for response interpretation, then discuss findings about user behavior,
% and then investigate the contributions of the system components 
and finally, investigate the factors that make the task challenging.

\begin{table*}[]
    \centering
    \caption{Questionwise Results: Accuracy is reported for the best performing model; Fleiss Kappa is agreement among human annotators; the last row is the fraction of times annotators chose "None of the above". Response length represents the number of words in the response}
    \label{tab: questionwise_results}
    \begin{tabular}{llrrrrrrrrrrrrrr}
&  Q1 &  Q2 &  Q3 &  Q4 &  Q5 &  Q6 &  Q7 &  Q8 &  Q9 &  Q10 &  Q11 &  Q12 & Mean \\
\hline
Accuracy & 0.76 & 0.66 & 0.76 & 0.81 & 0.60 & 0.71 & 0.58 & 0.52 & 0.40 & 0.41 & 0.84 & 0.69 & 0.65 \\
Fleiss $\kappa$ & 0.88 & 0.78 & 0.78 & 0.74 & 0.69 & 0.49 & 0.44 & 0.26 & 0.22 & 0.21 & 0.09 & 0.04 & 0.47 \\
Number of Options & 2.00 & 3.00 & 11.00 & 4.00 & 5.00 & 4.00 & 5.00 & 4.00 & 4.00 & 3.00 & 4.00 & 3.00 & 4.33 \\
% \cdashline{1-14}[0.5pt/5pt]
\hline
Conversational Dwell Time (sec) & 11.79 & 7.38 & 6.21 & 12.42 & 9.38 & 12.03 & 12.23 & 14.77 & 14.57 & 13.43 & 7.52 & 6.68 & 10.70 \\
Conversational Response Length & 3.30 & 2.78 & 3.44 & 6.70 & 3.44 & 6.41 & 5.48 & 7.30 & 4.19 & 4.63 & 4.00 & 4.19 & 4.65 \\
% \cdashline{1-14}[0.5pt/5pt]
\hline
Questionnaire Dwell Time (sec) & 10.59 & 4.96 & 10.93 & 14.37 & 9.15 & 7.52 & 10.04 & 8.59 & 20.70 & 20.56 & 7.74 & 11.30 & 11.37 \\
Questionnaire Response Length & 1.23 & 1.95 & 3.65 & 1.40 & 7.71 & 4.26 & 7.46 & 2.71 & 5.04 & 2.73 & 4.88 & 1.42 & 3.70 \\
% \cdashline{1-14}[0.5pt/5pt]
\hline
"None of the above" answers & 0.02 & 0.03 & 0.08 & 0.08 & 0.14 & 0.41 & 0.22 & 0.58 & 0.37 & 0.50 & 0.02 & 0.64 & 0.26 \\
    \end{tabular}
\end{table*}

\subsection{Response Interpretation Results}
Table \ref{tab:main_results_accuracy} shows the accuracy of all the models on single-option questions. We consider improvement to be statistically significant if ttest on each fold returns a p-value < 0.05. Significant results are marked in bold text.

The accuracy of TOD-BERT is not significantly higher than our probabilistic baseline. 
This is because the conversations in our setting are different from the goal-oriented dialog that TOD-BERT is pre-trained on. The model is not able to transfer its knowledge to response interpretation in a structured interview. 

Fine-tuned BERT and BERT-CNNET significantly outperform the baseline. 

The highest value of accuracy we achieve is 64\%, which is relatively low. 
As discussed in \S \ref{sec: agreement}, the inter-annotator agreement is  lower on some questions, indicating that intent prediction on these questions is difficult even for humans. 
We obtain higher accuracy values by excluding questions with low inter-annotator agreement from our test set. 
We set the threshold for low agreement as 0.4, which is standard for Fleiss Kappa. This leaves us with 7 single-option questions out of 12. 
Table \ref{tab:main_results_accuracy} also shows these results.
% maybe add back later
% We compare 4 versions of our models, one that uses just the current chat response as the conversation representation, one that uses the current question and its response, one that includes 1 previous turn of conversation in addition to the current, and lastly, on that includes 2 previous turns of conversation in addition to the current. 
% Unfortunately, incorporating previous turns of conversation does not improve accuracy (decreases by $\sim4\%$ on average). This might be because each information turn asks a new question. A structured interview flows differently than an ordinary conversation. 
% Results are reported for the version that uses the current question and its response as conversation representation.

\subsection{Tradeoff Between Effort and Information}
Table \ref{tab: questionwise_results} summarizes our findings from the user study.
The average dwell time (Time elapsed between the question's appearance and the user's first click/keypress) for a question was comparable for \FORM and \CHAT. The input time was much longer for \CHAT because participants had to type their responses instead of selecting options with clicks.
On average, the \CHAT response has more words than the \FORM response.
In some cases, the extra effort on the users' part resulted in more informative answers. 
For example, for the questions, "When do you moisturize your face"? (Q4) and "How do you handle unexpected stress?" (Q8), the \CHAT response is significantly more verbose than the \FORM response.
These questions elicited descriptive answers that were more informative in \CHAT.

On the other hand, for the question "What kind of hair day are you having today?" (Q5), users were more likely to give a response like "good" or "not bad". Although the longest conversational response for this question had 13 words, on average \FORM elicited more informative responses.

We also observe that 26\% of the \CHAT responses annotated by MTurkers were mapped to "None of the above", which indicates that \CHAT often collects information that is entirely absent from \FORM options. The highest number of "None of the above" responses were observed for questions "After applying a facial moisturizer, how do you like your skin to feel?" (Q10) and "How would you describe your
natural hair?" (Q12). This might have been because these questions can be interpreted in different ways, but the options list is small and specific. 

\subsection{Error Analysis}

Table \ref{tab:correlations} shows the correlation between 4 features of questions with the best model's accuracy (Accuracy) and the inter-annotator agreement ($\kappa$) for that question.
Contrary to what we expected, a larger number of options does not make the task harder for the model or human annotators. 
The number of words in the conversational response (Conv. Response Length) negatively correlates with $\kappa$ more than with Accuracy. That might be because longer responses could partially match more than one answer option and cause disagreement.
A longer dwell time indicates that the question is hard to understand or hard to answer. It negatively correlates with Accuracy more than with $\kappa$. This might be because it is harder for the model to handle unusual responses it hasn't been trained on.

Thus, we can see that the model fails to generalize to unusual responses. Another case where we observe high error is when matching responses requires some logical reasoning.
For example, for the question "Which ONE benefit are you primarily looking for, over time, from your facial moisturizer products?", the user responds by saying "The main benefit I'm looking for is smooth/healthy looking skin that isn't oily or shiny". However, the choices in the questionnaire are "Maintain the appearance/feel of my skin", "Enhance my skin's appearance/ feel", "Fix my skin's problem areas" and "Prevent future skin problems". The model would have to infer that the user's response implies that they want to enhance their skin's appearance. 
The domain-specific nature of the task also remains a source of error. ConceptNet does not have high enough coverage of skincare terms.

\section{Conclusion and Future Work}
In summary, we conducted a study to investigate the difference in responses between \CHAT and \FORM.
We find that \CHAT has the advantage of eliciting an answer that might not be one of the options but is informative of the user's preferences.
We also see that \CHAT elicits descriptive, more informative answers from users for open-ended questions.
% but also makes automated response-interpretation harder.
On the other hand, questions that ask for specific information and have a comprehensive list of options can be answered more efficiently using \FORM.

Moreover, we investigated the problem of automated response interpretation in a conversational structured interview setting, which is more challenging than the traditional intent classification task. 
We compared three complementary approaches to this problem, namely incorporating historical information, conversation context, and external knowledge for more effective semantic matching, all using state-of-the-art contextual large language models to represent the conversational and structured data. 
Our results demonstrate that effectively incorporating contextual information in structured interviews is harder than in other types of dialog. Although responses to previous interview questions can contain clues to infer future responses, we could not capture them by concatenating previous turns with the input to our model.
A possible future research direction would be to create a more effective context representation for structured interviews.
Another direction of research we plan to pursue is automatically adapting the conversation to ask clarification questions if the participants' response is unclear or to even skip some questions if the participant already provided information matching one of the options. 
Such an adaptive system can also use a combination of open-ended conversational interaction and suggesting options when necessary. 
Lastly, incorporating external knowledge in the absence of an appropriate knowledge graph, possibly using unstructured text from our domain, is another direction we plan to explore.

    % \section{Acknowledgment}
    
\begin{acks}
This research was supported by Procter \& Gamble.
\end{acks}

\begin{table}[h]
    \centering
    \begin{tabular}{lrrrr}
    & \multicolumn{2}{c}{Pearson}   & \multicolumn{2}{c}{Spearman} \\
    & Accuracy & $\kappa$               & Accuracy & $\kappa$                \\
    \hline
    % accuracy 7 default & 1.0 & 0.37 & 1.0 & 0.29 \\
    % k & 0.37 & 1.0 & 0.29 & 1.0 \\
    % kramers k & 0.37 & 1.0 & 0.29 & 1.0 \\
    No. of Options & 0.18 & 0.26 & 0.05 & 0.04 \\
    Conv. Response Length & -0.1 & -0.24 & -0.21 & -0.42 \\
    % chat input & -0.34 & -0.56 & -0.29 & -0.49 \\
    % chat total & -0.53 & -0.40 & -0.58 & -0.43 \\
    Dwell Time (Conversational) & -0.61 & -0.14 & -0.59 & -0.21 \\
    Dwell Time (Online Survey) & -0.58 & -0.30 & -0.27 & -0.25 \\
    % "None of the Above" Count & -0.58 & -0.69 & -0.62 & -0.60 \\
    % form total & -0.63 & -0.34 & -0.31 & -0.24 \\
    % form input & -0.72 & -0.51 & -0.71 & -0.58 \\
    \end{tabular}
    \caption{Correlation Values}
    \label{tab:correlations}
\end{table}

\bibliographystyle{ACM-Reference-Format}
\balance
\bibliography{sample-base}

\end{document}